\documentclass[a4paper,11pt]{article}
\usepackage{pos}
\newcommand{\CS}{\textcolor{blue!60!black}}
\newcommand{\nf}{n_f}
\newcommand{\ud}{\mathrm{d}}

\graphicspath{ {figures/} }
\newcommand{\FDiag}[2]{
\begin{minipage}{0.18\textwidth}
\begin{center}
\CS{#1}\\[-2ex]
\includegraphics[angle=-90,width=\textwidth]{#2}
\end{center}
\end{minipage}
\hspace*{-2ex}
}

\title{
\vspace*{-2ex}
{\hfill
\rm\footnotesize ZU-TH 33/26}
\vspace*{2ex}\\
Non-singlet splitting functions at four loops
}
\author[a]{Thomas Gehrmann}
\author[b]{Andreas von Manteuffel}
\author[a]{Vasily Sotnikov}
\author[c,d]{Tong-Zhi~Yang}

\affiliation[a]{Physik-Institut, Universit\"at Z\"urich, Winterthurerstrasse 190, 8057 Z\"urich, Switzerland}
\affiliation[b]{Institut f\"ur Theoretische Physik, Universit\"at Regensburg, 93040 Regensburg, Germany}
\affiliation[c]{State Key Laboratory of Nuclear Physics and Technology, Institute of Quantum Matter,
South China Normal University, Guangzhou 510006, China}
\affiliation[d]{%
Guangdong Basic Research Center of Excellence for Structure and Fundamental Interactions of Matter, Guangdong Provincial Key Laboratory of Nuclear Science, Guangzhou 510006, China
}

\emailAdd{thomas.gehrmann@uzh.ch}
\emailAdd{manteuffel@ur.de}
\emailAdd{vasily.sotnikov@physik.uzh.ch}
\emailAdd{tongzhi.yang@m.scnu.edu.cn}

\abstract{We present a complete calculation of the non-singlet splitting functions
in four-loop quantum chromodynamics.
}

\FullConference{Based on talk presented by Andreas von Manteuffel at\\
Loops and Legs in Quantum Field Theory (LL2026),\\
12-17, April, 2026, 
Bayreuth, Germany.\\[2ex]
Based on talk presented by Vasily Sotnikov at\\33rd International Workshop on Deep Inelastic Scattering and Related Subjects (DIS2026),\\
4-8, May, 2026,
Bologna, Italy.
}

\allowdisplaybreaks[1]

\begin{document}
\maketitle

\section{Introduction}
Parton distribution functions lie at the heart of standard collinear
factorization for hadronic scattering processes.
To describe the scale evolution of parton distribution functions,
one considers the gluon distribution $g(x,\mu^2)$ and suitably formed combinations
of quark densities $q_i(x,\mu^2)$ and anti-quark densities $\bar{q}_i(x,\mu^2)$,
where $x$ is the momentum fraction of the parton, $\mu$ the factorization scale,
and $i=1,\ldots,\nf$ denotes a massless flavor.
Specifically, one defines flavor asymmetries $q_{\text{ns},ij}^+$ and $q_{\text{ns},ij}^-$,
the flavor non-singlet valence distribution $q_{\text{ns}}^\text{v}$
and the flavor singlet distribution $q_{\text{s}}$ according to
\begin{align}
    q_{\text{ns},ij}^\pm = (q_i\pm\bar{q}_i)-(q_j\pm\bar{q}_j),\qquad
    q_{\text{ns}}^\text{v} = \sum_i (q_i-\bar{q}_i),\qquad
    q_\text{s} = \sum_i (q_i + \bar{q}_i)\,.
\end{align}
This decomposition decouples the Dokshitzer-Gribov-Lipatov-Altarelli-Parisi
scale evolution for the non-singlet distributions,
\begin{align}
\frac{\ud}{\ud \ln \mu^2} q_{\text{ns},ij}^a &=
  P_\text{ns}^a \otimes q_{\text{ns},ij}^a\,,\quad a=+,-,\text{v},
\end{align}
while the evolution of singlet quark and gluon distribution is described by a coupled $2\times 2$ system,
\begin{align}
\frac{\ud}{\ud \ln \mu^2} 
\begin{pmatrix} q_\text{s} \\ g \end{pmatrix}
    &= \begin{pmatrix} P_{qq} & P_{qg} \\
    P_{gq} & P_{gg} \end{pmatrix} \otimes 
    \begin{pmatrix} q_\text{s} \\ g \end{pmatrix}.
\end{align}
Here, the circled cross denotes Mellin convolution,
\begin{equation}
    (f \otimes g)(x)
    \equiv \int_0^1 \ud y \int_0^1 \ud z\,
        \delta(x-yz)f(y)g(z).
\end{equation}
The splitting functions $P^+_\text{ns}(x,\mu^2)$, $P^-_\text{ns}(x,\mu^2)$, $P^\text{v}_\text{ns}(x,\mu^2)$, $P_{qq}(x,\mu^2)$, $P_{qg}(x,\mu^2)$, $P_{gq}(x,\mu^2)$, $P_{gg}(x,\mu^2)$ can be calculated perturbatively for $\nf$ massless quark flavors.
Their exact 3-loop expressions have been known since some time now~\cite{Moch:2004pa,Vogt:2004mw,Blumlein:2021enk,Blumlein:2021ryt,Gehrmann:2023ksf}.
We note in passing that also heavy flavor corrections to the evolution of structure functions have been pushed to the 3-loop level more recently, see \cite{Ablinger:2026xza} for a review in these proceedings.
Fully consistent N${}^3$LO predictions of cross sections at the Large Hadron Collider require 4-loop splitting functions, for which only numerical approximations \cite{Moch:2017uml,Moch:2021qrk,Falcioni:2023luc,Falcioni:2023vqq,Falcioni:2024xyt,Falcioni:2024qpd,Falcioni:2025hfz} and partial analytical results \cite{Gracey:1994nn,Davies:2016jie,Gehrmann:2023cqm,Gehrmann:2023iah,Falcioni:2023tzp,Kniehl:2025ttz,Kniehl:2025jfs} were available previously.

In this proceedings contribution, we report on a complete calculation of the non-singlet splitting functions $P^+_\text{ns}$, $P^-_\text{ns}$, and $P_\text{ns}^s = P_\text{ns}^\text{v}- P_\text{ns}^-$ at four loops~\cite{Gehrmann:2026qbl}.
We gratefully acknowledge reference~\cite{Moch:2026qsw}, in which our results were checked against further independent fixed-moment calculations, shown to be structurally consistent with theoretical expectations, and used to derive further quantities of interest.

\section{Calculational method}

The Mellin transformations of the non-singlet splitting functions
\begin{align}
    \gamma_\text{ns}^{\pm,\text{s}}(n,\mu) &= -\int_0^1\ud x\, x^{n-1} P_\text{ns}^{\pm,\text{s}}(x,\mu)
\end{align}
for integer Mellin moment $n$ appear as the anomalous dimensions of twist-two, spin $n$ operators
\begin{equation}
O_{\mathrm{ns}}(n) = \frac{i^{n-1}}{2}  \bar{\psi} \,\Delta\cdot \gamma \,(\Delta \cdot D)^{n-1}\,\frac{\lambda}{2} \psi  
\end{equation}
where $\Delta$ is a light-like auxiliary momentum and $\lambda$ a matrix in flavor space used to project onto different non-singlet quark combinations.
The renormalization of these operators takes the simple form
\begin{equation}
    O^{\text{R}}_{\text{ns}}(n,\mu^2)  = Z_{\text{ns}}(n,\mu^2) \,O^{\text{B}}_{\text{ns}}(n)\,,
\end{equation}
and the anomalous dimensions can be found from
\begin{equation}
\label{eq:anomdimdef}
 \frac{\ud}{\ud \ln \mu^2} Z_{\text{ns}}(n,\mu^2) = -  \gamma_{\text{ns}}(n,\mu^2) \, Z_{\text{ns}}(n,\mu^2)
\end{equation}
once $Z_{\text{ns}}(n,\mu^2)$ is known to the required loop order.
The latter can be computed by evaluating operator matrix elements $\langle q \vert O_{\mathrm{ns}}(n) \vert q \rangle$ with two off-shell quarks.
We use dimensional regularization with the regulator $\epsilon=(4-d)/2$.

\begin{figure}[t]
\centering
 \FDiag{$C_A^3 C_F $}{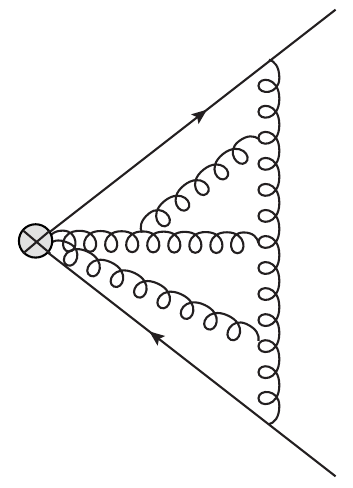}\;\;
 \FDiag{$C_A^2 C_F^2 $}{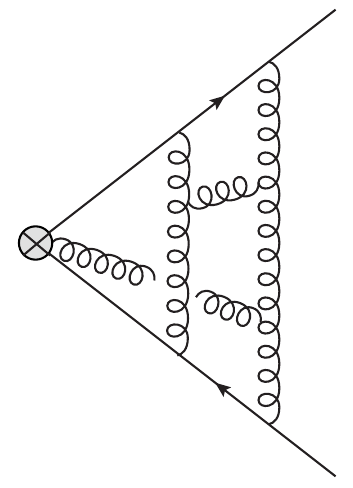}\;\;
 \FDiag{$C_A C_F^3 $}{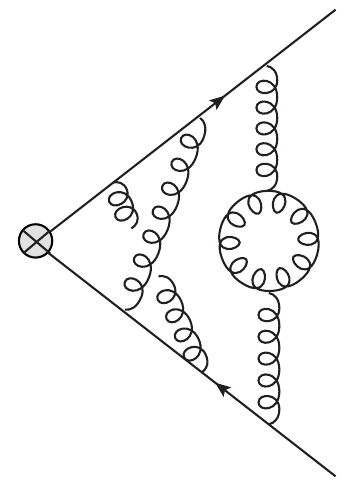}\;\;
 \FDiag{$C_F^4$}{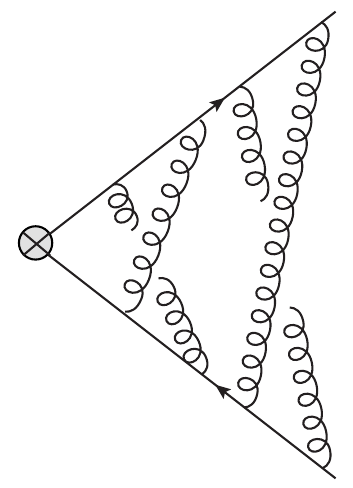}\;\;
 \FDiag{$d_A^{abcd}d_F^{abcd}/N_c$}{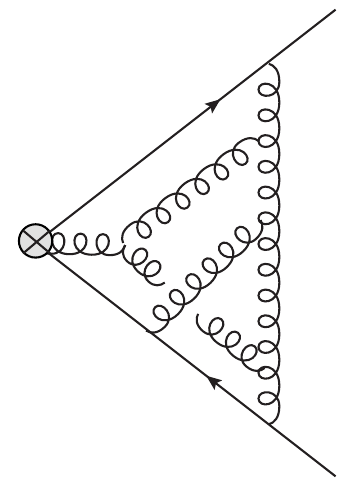}
\\[1ex]
 \FDiag{$\nf (d_F^{abc})^2 C_A / N_c$}{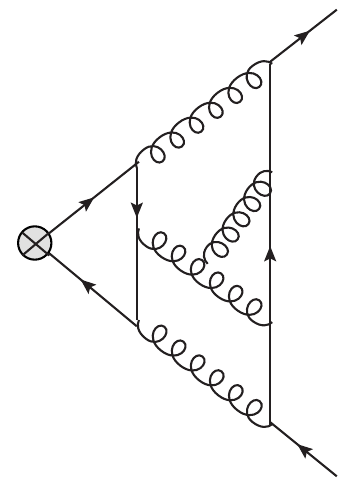}\;\;
 \FDiag{$\nf (d_F^{abc})^2 C_F / N_c$}{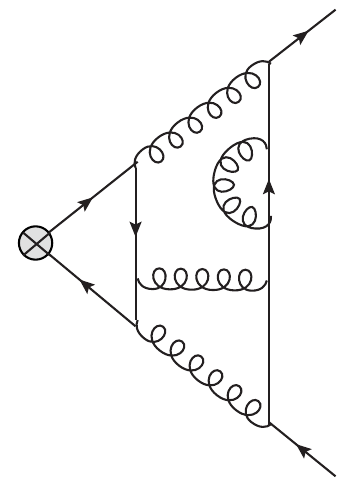}
\caption{\label{fig:diags}
Four-loop Feynman diagrams for operator matrix elements with off-shell external quarks.
They contribute to different color coefficients of the splitting functions 
$P_\mathrm{ns}^{(3) \pm}$ (upper row) and $P_\mathrm{ns}^{(3) s}$ (lower row).
Further color structures arise from diagrams with additional closed quark loops.
}
\end{figure}
Figure~\ref{fig:diags} shows example diagrams contributing to the four-loop color coefficients computed fully analytically in \cite{Gehrmann:2026qbl} for the first time for arbitrary values of $n$.
We employ \texttt{Qgraf}~\cite{Nogueira:1991ex}, \texttt{Form}~\cite{Vermaseren:2000nd,Davies:2026cci}, \texttt{Color.h}~\cite{vanRitbergen:1998pn},
\texttt{Reduze\;2}~\cite{vonManteuffel:2012np} and the private integration-by-parts code \texttt{Finred} based on finite field sampling~\cite{vonManteuffel:2014ixa,Peraro:2016wsq} to compute the matrix elements.
The operator insertion leads to $n$-dependent powers of scalar products.
In order to keep $n$ arbitrary and still be able to use standard integration-by-parts identities, we make use of generating functions~\cite{Ablinger:2012qm,Ablinger:2014nga}.
We replace
 \begin{align}
&(\Delta \cdot p)^{n-1} \to \sum_{n=1}^\infty t^n (\Delta \cdot p)^{n-1} = \frac{t}{1-t \Delta \cdot p } \,, \nonumber \\
 &  \sum^{n-3}_{j=0} (\Delta \cdot p_1)^{n-3-j} (\Delta \cdot p_2)^{j} \to \sum^{\infty}_{n=3} t^n \sum^{n-3}_{j=0} (\Delta \cdot p_1)^{n-3-j} (\Delta \cdot p_2)^{j} = \frac{t^3}{(1-t \Delta\cdot p_1)(1- t \Delta \cdot p_2)} \,, \nonumber \\
 &\text{etc.,}
 \end{align}
which introduces linear propagators in addition to the standard quadratic ones in the Feynman integrals.
This allows us to express the matrix elements in terms of master integrals, which depend on $t$.

\begin{figure}[t]
  \centering
  \includegraphics[width=0.27\linewidth]{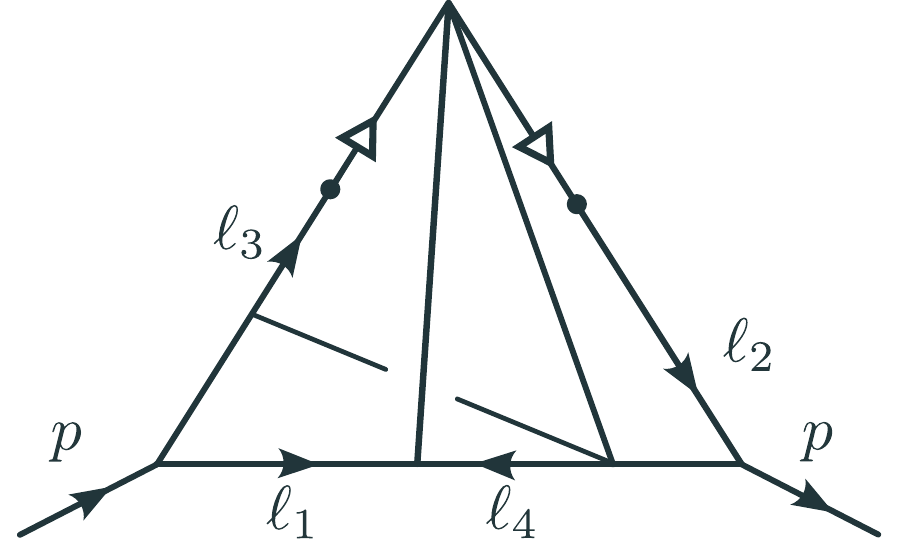}
  \caption{  \label{fig:elliptic}
  An elliptic sector contributing to the fifth Feynman diagram of figure~\ref{fig:diags}.
  In addition to nine standard propagators, there are two denominators $1- t\,(\Delta \cdot \ell_i)$, $i=1,2$, involving the tracing parameter $t$.}
\end{figure}

To calculate the master integrals as a Laurent series in $\epsilon$, we derive differential equations in the tracing parameter $t$.
The general solutions are quite involved.
For one of the simpler color factors~\cite{Gehrmann:2023iah}, we determined a canonical basis and solved the $\epsilon$-factorized differential equations in terms of multiple polylogarithms of $t$.
For the complete set of contributions, this is a more challenging task and involves also non-trivial geometries.
For example, using the package \texttt{DLogBasis} \cite{Henn:2020lye}, we find that the topology shown in figure~\ref{fig:elliptic} is related to an elliptic curve on the maximal cut.

We find it convenient to circumvent these challenges by solving the differential equations only for an expansion about $t=0$.
For our purposes, it is sufficient to consider Taylor expansions in $t$, which are regular at $t=0$.
For $t=0$ the Feynman integrals take the form of self-energy integrals with only standard quadratic propagators, such that the known four-loop propagator master integrals can be used as boundary information.
The usage of recursion relations~\cite{Blumlein:2009tj, Ablinger:2015tua, Lee:2017qql} allows us to compute the master integrals and from this the matrix elements up to high orders in $t$.
In this way, we compute the anomalous dimensions for $n$ up to about $4$k.

Having computed this large amount of Mellin moments allows us to infer the solution for arbitrary $n$.
We make an ansatz in terms of harmonic sums
\begin{align}
S_{\pm m_1, \,m_2,\,\cdots m_d}(n) = \sum_{j=1}^{n} (\pm 1)^{j} j^{-m_1} S_{m_2,\,\cdots m_d}(j), \quad m_i \in \mathbb{N},\qquad
S_\emptyset(n)=1\,,
\end{align}
up to weight 7, Riemann zeta values up to weight 5, and denominator factors 
    $n$, $n \pm 1$, $n \pm 2$. 
By comparing the ansatz to our fixed Moment results, we were able to identify the all-$n$ result.

While our construction assumed specific building blocks for the anomalous dimensions, this is not strictly necessary.
Johannes Bl\"umlein and Carsten Schneider reported~\cite{Bluemlein:2026cf4recursion} that the non-zeta term in the $C_F^4$ part of our result for $\gamma_\text{ns}^+$ can be reconstructed from fixed Mellin moments without structural assumptions such as guessing denominators and assuming harmonic sums. They were able to determine a recursion of order 28 and degree 1015 from 5520 even moments generated from our expressions, from which the full result could have been reconstructed.
This corresponds to about $11$k terms in the $t$ expansion, which is more than what was needed in our approach but still computationally feasible.

\section{Results}

While the complete results are too lengthy to be displayed here, we list a few terms for illustrative purposes in the following. For example, we find
\begin{align}
    \gamma_{\mathrm{ns}}^{(3),+} &=
     \tfrac{1}{2}(1 + (-1)^n) \big(
     \CS{C_A^3 C_F}\big[
     d_0^6 ( -912 S_1 )
     +d_1^6 ( -912 S_1 )
     +d_0^5 ( 1024/3  S_{1,1} + 608 S_{-2} - 512/3  S_{2} 
     \nonumber\\ &
     + 640 S_{1} )
     +d_1^5 ( - 1024/3 S_{1,1} - 608 S_{-2} + 512/3 S_{2} - 2400 S_{1} )
     + d_0^4 (3296/3 S_{-2, 1}
     \nonumber\\ &
     - 944/3 S_{1,-2} - 392 S_{-3} + 616/3 S_3 - 3664/3 S_{1,1}   - 2512/3 S_{-2}  + 1172/3 S_2 - 1732/3 S_1
     \nonumber\\ &
     + 2662/27 )
     + d_1^4 (3296/3 S_{-2, 1} - 944/3 S_{1, -2} - 392 S_{-3} + 616/3 S_3 - 3664/3 S_{1, 1}
     \nonumber\\ &
     - 3040/3 S_{-2} + 1172/3 S_2 - 7804/3 S_1  - 2662/27 )
+ \ldots
        \big]
        + \ldots \big)
\end{align}
for the perturbative expansion $\gamma_\text{ns}^+ = \sum_{\ell=1}^\infty a_s^\ell \gamma_\text{ns}^{(\ell - 1),+}$ with $a_s \equiv \alpha_s(\mu^2)/(4\pi)$.
Here, we use the abbreviations $d_0=1/n$, $d_1=1/(n+1)$ and suppress the argument $n$ of the harmonic sums.
For $\gamma_\text{ns}^{(3),-}$, similar structures appear, but in this case there are contributions only for odd instead of even $n$.
In both anomalous dimensions, $n$ and $n+1$ are the only denominator factors.
For $\gamma_\text{ns}^{(3),\text{s}}$, non-zero contributions arise for odd $n$ and involve denominator factors $n$, $n\pm 1$, $n+2$, but not $n-2$.
We confirm the all-$n$ results for all known color coefficients~\cite{Gracey:1994nn,Davies:2016jie,Moch:2017uml,Gehrmann:2023iah,Kniehl:2025ttz}.
For the new color coefficients, our results reproduce known lower Mellin moments~\cite{Moch:2017uml} and agree with further moments~\cite{Moch:2026qsw} and predictions~\cite{Kniehl:2026axe} which were not available at the time of our first arXiv submission.

\begin{figure}
\centering
\includegraphics[width=0.6\textwidth]{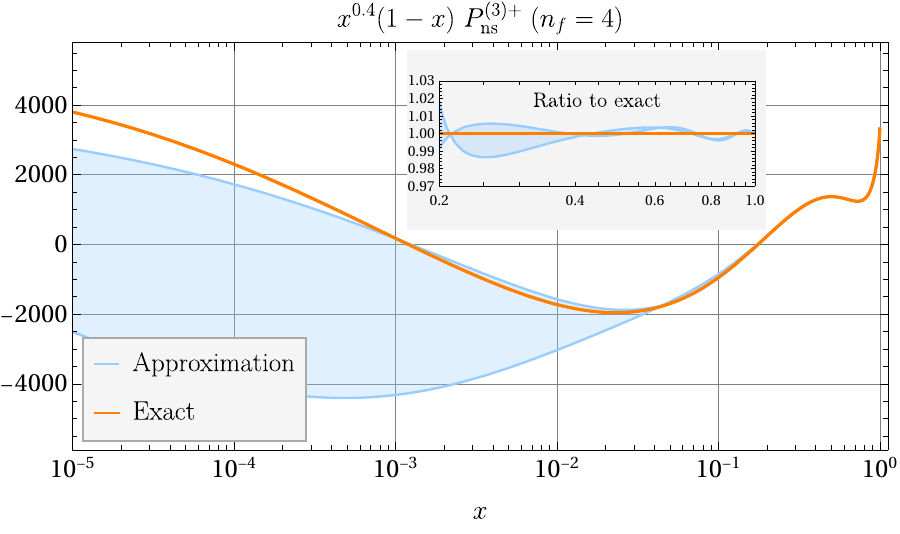}
\caption{\label{fig:splittingplus}
Non-singlet, four-loop splitting function $P_\text{ns}^{(3),+}(x)$ \cite{Gehrmann:2026qbl} multiplied by $x^{0.4}(1-x)$ to enhance readability.
We compare our exact result to the numerical approximation of reference~\cite{Moch:2017uml}.
}
\end{figure}

We obtain the splitting functions in $x$ space through inverse Mellin transform with the help of the package \texttt{HarmonicSums}~\cite{Ablinger:2009ovq,Ablinger:2012ufz}.
The results for $P_\text{ns}^{(3),\pm}$ contain harmonic polylogarithms up to weight six, zeta values up to weight seven, the delta distribution $\delta(1-x)$, the plus distribution $[1/(1-x)]_+$, denominator factors $1-x$ and $1+x$, and positive powers of $x$.
For $P_\text{ns}^{(3),\text{s}}$ we find harmonic polylogarithms and zeta values up to weight six, denominator factors $x$, $1-x$ and $1+x$, and positive powers of $x$.
For numerical evaluations of the harmonic polylogarithms, we employ the package \texttt{HPL} \cite{Maitre:2005uu}
and the implementation \cite{Vollinga:2004sn} of $G$ functions in \texttt{GiNaC}.
Overall, our complete results agree well with the numerical approximations of \cite{Moch:2017uml} given the quoted error bands.
For $P_\text{ns}^{(3),+}$ at small values of $x$, our result lies a bit above the error band, as can be seen from figure~\ref{fig:splittingplus}.
Our results help to reduce the theoretical error for the non-singlet splitting functions, in particular for small values of $x$.

In the small $x$ limit, $x\to 0$, the splitting functions diverge logarithmically, with the leading logarithmic contribution being proportional to $\ln^6 x$.
For $P_\text{ns}^{(3),+}$, we can compare our result against the prediction of reference~\cite{Davies:2022ofz}.
Our result reads
\begin{align}
    P_\text{ns}^{(3),+}  &\xrightarrow[x\to 0]{}
    \ln^6 x \,\biggl[ \CS{C_F^4} \frac{1}{9} \biggr]
    + \ln^5 x \,\biggl[ \CS{C_A C_F^3} \frac{22}{9}
      - \CS{C_F^4} \frac{4}{3}
      - \CS{C_F^3 \nf} \frac{4}{9} \biggr]
    + \ln^4 x \,\biggl[
      \CS{C_A^3 C_F} 38 \zeta_2
      \nonumber \\ & \qquad 
    + \CS{C_A^2 C_F^2} \biggl( -140 \zeta_2 + \frac{121}{9}\biggr)
    + \CS{C_A C_F^3} \biggl( 168 \zeta_2 + \frac{170}{9} \biggr)
    + \CS{C_F^4} \biggl( -\frac{248}{3}\zeta_2 +\frac{16}{3} \biggr)
      \nonumber \\ & \qquad 
    + \CS{\frac{d_A^{abcd}d_F^{abcd}}{N_c}} ( -48 \zeta_2 )
    + \CS{C_A C_F^2 \nf} \biggl( -\frac{44}{9} \biggr)
    + \CS{C_F^3 \nf} \biggl( -\frac{20}{9} \biggr)
    + \CS{C_F^2 \nf^2} \biggl( \frac{4}{9} \biggr)
    \biggr]
    + \ldots
\end{align}
While we find that the $\ln^6 x$ and $\ln^5 x$ terms coincide with \cite{Davies:2022ofz}, the $\ln^4 x$ terms deviate unexpectedly.
The difference of our $\ln^4 x$ terms to those predicted by \cite{Davies:2022ofz} is
\begin{align}
\left( \CS{C_A^3 C_F} 38 - \CS{C_A^2 C_F^2} 120 
+ \CS{C_A C_F^3} 104 - \CS{C_F^4} 16 - \CS{\frac{d_A^{abcd}d_F^{abcd}}{N_c}} 48 \right) \zeta_2 \,.
\end{align}
Replacing all Casimir operators by their $SU(N_c)$ values the difference is
\begin{align}
    \frac{ (N_c^2-1)( N_c^4 + 10 N_c^2 + 1)}{N_c^4} \zeta_2\,.
\end{align}
and thus subleading in $N_c$.
It would be curious to investigate the reason for this deviation and whether it is, for example, due to a non-trivial phase inducing $(i\pi)^2$ terms.

In the limit $x \to 1$, the splitting function $P_\text{ns}^{(3),s}$ vanishes, while
$P_\text{ns}^{(3),+}$ and $P_\text{ns}^{(3),-}$  
become equal up to subleading terms.
We can decompose
\begin{align}
P_\text{ns}^{(3),+} &\xrightarrow[x\to 1]{}
A^{(4)}_q \left[\frac{1}{1-x}\right]_+
+  B_q^{(4)} \, \delta(1-x)
+  C^{(4)}_q \log (1-x)+  D^{(4)}_q - A^{(4)}_q + \mathcal{O}(1-x)\,.
\end{align}
Our calculation reproduces the known four-loop coefficient $A_q^{(4)}$~\cite{Henn:2019swt,vonManteuffel:2020vjv} of the cusp anomalous dimension \cite{Korchemsky:1987wg}.
For the so-called virtual anomalous dimension $B_q^{(4)}$ we determine the fully analytical result
\begin{align}
 B_q^{(4)} &=  \CS{C_A^3 C_F} \biggl( 
  - \frac{8960 }{3}\zeta_7
  + \frac{1472}{3} \zeta_5 \zeta_2
  + \frac{32}{3} \zeta_4 \zeta_3
  + \frac{73333}{108} \zeta_6
  + \frac{1672}{3} \zeta_3^2
  + \frac{11522 }{9}\zeta_5
  + \frac{584}{3} \zeta_3 \zeta_2
\nonumber \\ &
  - \frac{11206}{27} \zeta_4
  - \frac{152284 }{81}\zeta_3
  + \frac{13864}{9} \zeta_2
  - \frac{373793}{648}
  \biggr)
  + \CS{C_A^2 C_F^2} \biggl( 
   8610 \zeta_7
  - 2104 \zeta_5 \zeta_2
  - 32 \zeta_4 \zeta_3
\nonumber \\ &
  - \frac{5497}{2}\zeta_6
  - \frac{7102 }{3}\zeta_3^2
  + \frac{5354}{9} \zeta_5
  + \frac{2096 }{9}\zeta_3 \zeta_2
  - \frac{60850 }{27}\zeta_4
  + \frac{129662 }{27}\zeta_3
  - \frac{46771 }{27}\zeta_2
  + \frac{29639}{36}
  \biggr) \nonumber \\
&+ \CS{C_A C_F^3} \biggl( 
  - 10920 \zeta_7
  + 2064 \zeta_5 \zeta_2
  + 128 \zeta_4 \zeta_3
  + \frac{79297}{18}\zeta_6
  + 3220 \zeta_3^2
  - 976 \zeta_5
  - \frac{1988 }{3}\zeta_3 \zeta_2
\nonumber \\ &
  + 2167\zeta_4
  - 3260 \zeta_3 
  + 1167 \zeta_2
  - \frac{2085}{4}
  \biggr)
  + \CS{C_F^4} \biggl( 
   5880 \zeta_7
  - 384 \zeta_5 \zeta_2
  + 64 \zeta_4 \zeta_3
  -2111 \zeta_6
\nonumber \\ &
  - 1152 \zeta_3^2
  - 2520 \zeta_5
  - 120 \zeta_3 \zeta_2
  - 342 \zeta_4
  + 2004 \zeta_3
  - 450 \zeta_2
  + \frac{4873}{24}
  \biggr)
  + \CS{\frac{d_F^{abcd} d_A^{abcd} }{N_c} } \biggl( 
   2800 \zeta_7
\nonumber \\ &
  + 320 \zeta_5 \zeta_2
  - 64 \zeta_4 \zeta_3
  -\frac{1562}{9}\zeta_6
  - 704 \zeta_3^2
  - 400 \zeta_5 
  - 896 \zeta_3 \zeta_2
  + \frac{32}{3} \zeta_4
  - \frac{1232 }{3}\zeta_3
  - \frac{944 }{3}\zeta_2
  + 96 
  \biggr)
\nonumber \\ &
+ \CS{C_A^2 C_F \nf} \biggl( 
  -\frac{3913}{27} \zeta_6
  + \frac{416 }{3}\zeta_3^2
  + \frac{1130 }{9}\zeta_5
  - \frac{580 }{3}\zeta_3 \zeta_2
  + \frac{1234 }{9}\zeta_4
  + \frac{9554 }{27}\zeta_3
  - \frac{41092 }{81}\zeta_2
  \nonumber \\ &
  + \frac{20027}{108}
  \biggr)
  + \CS{C_A C_F^2 \nf} \biggl( 
   351\zeta_6
  - \frac{1232 }{3}\zeta_3^2
  - \frac{7432 }{9}\zeta_5
  + \frac{2672 }{9}\zeta_3 \zeta_2
  + \frac{27854}{27}\zeta_4
  \nonumber \\ &
  - \frac{15400 }{27}\zeta_3
  - \frac{3892 }{27}\zeta_2
  - \frac{7751}{54}
  \biggr)
  + \CS{C_F^3 \nf} \biggl( 
  -\frac{6434 }{9}\zeta_6
  + 224 \zeta_3^2
  + 912 \zeta_5
  - \frac{256 }{3}\zeta_3 \zeta_2
  - 204 \zeta_4
  \nonumber \\ &
  - 308 \zeta_3
  + 162 \zeta_2
  + 32
  \biggr)
  + \CS{\frac{(d_F^{abcd})^2 \, \nf}{N_c}} \biggl( 
   \frac{1792}{9}\zeta_6
  + 256 \zeta_3^2
  - 1120 \zeta_5
  + 64 \zeta_3 \zeta_2 
  - \frac{352}{3} \zeta_4
  \nonumber \\ &
  - \frac{992}{3} \zeta_3
  + \frac{1888}{3} \zeta_2
  - 192
  \biggr)
  + \CS{C_A C_F \nf^2} \biggl( 
  - \frac{88 }{9}\zeta_5
  + \frac{80 }{3}\zeta_3 \zeta_2
  -\frac{80}{9}\zeta_4
  - \frac{320 }{9}\zeta_3
  + \frac{3170 }{81}\zeta_2
  - \frac{193}{54}
  \biggr) \nonumber \\
&+ \CS{C_F^2 \nf^2} \biggl( 
   \frac{368 }{9}\zeta_5
  - \frac{160 }{9}\zeta_3 \zeta_2
  -\frac{2104 }{27}\zeta_4
  + \frac{56 }{27}\zeta_3
  + \frac{1244 }{27}\zeta_2
  - \frac{188}{27}
  \biggr)
  + \CS{C_F \nf^3} \biggl( 
  -\frac{32}{27}\zeta_4
  + \frac{304 }{81}\zeta_3
  \nonumber \\ &
  + \frac{32 }{81}\zeta_2
  - \frac{131}{81}
  \biggr)\,.
\end{align}

The virtual anomalous dimension is related to the soft anomalous dimension, which enters for example Drell-Yan threshold resummation.
In contrast to the virtual anomalous dimension, the soft anomalous dimension is an eikonal quantity \cite{Ravindran:2004mb,Idilbi:2006dg,Dixon:2008gr,Falcioni:2019nxk} and can thus be converted from the fundamental to the adjoint representation by a simple generalized Casimir rescaling.
The soft anomalous dimension is
\begin{align}\label{eq:softad}
    f_r = \gamma_r - 2 B_r
\end{align}
for quarks ($r=q$) and gluons ($r=g$), where $\gamma_r$ is the collinear anomalous dimension controlling the $1/\epsilon$ poles of $q\bar{q}\gamma$ and $ggH$ form factors.
Using the perturbative expansion
$f_r = \sum_{\ell=1}^\infty a_s^\ell f_r^{(\ell)}$,
$B_r = \sum_{\ell=1}^\infty a_s^\ell B_r^{(\ell)}$,
$\gamma_r = \sum_{\ell=1}^\infty a_s^\ell \gamma_r^{(\ell-1)}$,
such that 
$f_r^{(\ell)} = \gamma^{(\ell-1)}_r - 2 B_r^{(\ell)}$,
the four-loop results for the collinear anomalous dimension~\cite{Agarwal:2021zft,vonManteuffel:2020vjv} and our new four-loop result for the virtual anomalous dimension, we determine $f_q^{(4)}$.
Having the color algebra for a Wilson line based calculation in mind, we replace Casimir operators in the fundamental representation with Casimir operators in a general representation $R$, where $R=F$ for $r=q$ and $R=A$ for $r=g$.
We find
\begin{align}
    f_r^{(4)} &=
\CS{C_R C_A^3}\biggl(
-\frac{9671}{6}\zeta_7
-432\zeta_5 \zeta_2
-710\zeta_4 \zeta_3
-\frac{131197}{54}\zeta_6
-\frac{4906}{9}\zeta_3^2
+\frac{106034}{27}\zeta_5
+\frac{11896}{9}\zeta_3 \zeta_2
\nonumber \\ &
+\frac{115783}{27}\zeta_4
-\frac{837520}{243}\zeta_3
-\frac{1183819}{729}\zeta_2
+\frac{9364079}{6561}
\biggr)
+\CS{\frac{d_R^{abcd}d_A^{abcd}}{N_R}}\biggl(
-2116\zeta_7
+384\zeta_5 \zeta_2
\nonumber \\ &
-240\zeta_4 \zeta_3
+\frac{2200}{3}\zeta_6
+\frac{880}{3}\zeta_3^2
+\frac{5360}{9}\zeta_5
+16\zeta_4
-\frac{416}{9}\zeta_3
-96\zeta_2
\biggr)
+\CS{C_R C_A^2 \nf} \biggl(
\frac{16895}{27}\zeta_6
\nonumber \\ &
+\frac{4420}{9}\zeta_3^2
-\frac{692}{27}\zeta_5
-\frac{104}{9}\zeta_3 \zeta_2
-\frac{11050}{9}\zeta_4
-\frac{31340}{243}\zeta_3
+\frac{294539}{729}\zeta_2
-\frac{394109}{1944}
\biggr)
\nonumber \\ &
+\CS{C_R C_A C_F \nf}\biggl(
16\zeta_6
-312\zeta_3^2
+\frac{1448}{9}\zeta_5
-160\zeta_3 \zeta_2
-\frac{988}{9}\zeta_4
+\frac{68882}{81}\zeta_3
+\frac{2819}{9}\zeta_2
-\frac{813475}{972}
\biggr)
\nonumber \\ &
+\CS{C_R C_F^2 \nf}\biggl(
-200\zeta_6
-80\zeta_3^2
-\frac{1600}{3}\zeta_5
+74\zeta_4
+\frac{4424}{9}\zeta_3
-2\zeta_2
+\frac{21037}{108}
\biggr)
+\CS{C_R C_A \nf^2}\biggl(
-112\zeta_5
\nonumber \\ &
-\frac{224}{9}\zeta_3 \zeta_2
+\frac{388}{9}\zeta_4
+\frac{32152}{243}\zeta_3
-\frac{15481}{729}\zeta_2
+\frac{27875}{17496}
\biggr)
+\CS{C_R C_F \nf^2}\biggl(
\frac{304}{9}\zeta_5
+\frac{32}{3}\zeta_3 \zeta_2
+\frac{64}{9}\zeta_4
\nonumber \\ &
-\frac{4568}{81}\zeta_3
-\frac{172}{9}\zeta_2
+\frac{16733}{486}
\biggr)
+\CS{C_R \nf^3}\biggl(
\frac{128}{27}\zeta_4
-\frac{400}{243}\zeta_3
-\frac{16}{81}\zeta_2
-\frac{16160}{6561}
\biggr)
\nonumber \\ &
+\CS{\frac{d_R^{abcd}d_F^{abcd}\nf}{N_R}}\biggl(
-\frac{800}{3}\zeta_6
-\frac{320}{3}\zeta_3^2
-\frac{1600}{9}\zeta_5
-32\zeta_4
+\frac{640}{9}\zeta_3
+256\zeta_2
\biggr)
\end{align}
with $N_F=N_c$ and $N_A=N_c^2-1$ denoting the dimensionality of the representation $R$.

Employing equation~\eqref{eq:softad} for $r=g$ gives the prediction
\begin{align}
B_g^{(4)} &=
 \CS{C_A^4} \biggl(  \frac{1750}{3}\zeta_7
+\frac{200}{3}\zeta_5 \zeta_2
+\frac{512}{3}\zeta_4 \zeta_3
-\frac{37477}{108}\zeta_6
+\frac{770}{3}\zeta_3^2
-\frac{14467}{9}\zeta_5
-\frac{3566}{9}\zeta_3 \zeta_2
+\frac{8941}{54}\zeta_4
\nonumber \\ &
+\frac{48550}{27}\zeta_3
+\frac{2452}{27}\zeta_2
+\frac{48443}{486} 
  \biggr)
+\CS{\frac{(d_A^{abcd})^2}{N_A}}\biggl(  2800\zeta_7
+320\zeta_5 \zeta_2
-64\zeta_4 \zeta_3
-\frac{814}{9}\zeta_6
-704\zeta_3^2
\nonumber \\ &
-\frac{440}{3}\zeta_5
-1168\zeta_3 \zeta_2
-\frac{476}{3}\zeta_4
-672\zeta_3
+80\zeta_2
+\frac{64}{9} 
  \biggr)
  +\CS{C_A^3 \nf} \biggl(  \frac{1727}{54}\zeta_6
-\frac{836}{3}\zeta_3^2
+\frac{931}{3}\zeta_5
\nonumber \\ &
+\frac{718}{9}\zeta_3 \zeta_2
-\frac{1601}{54}\zeta_4
-\frac{12749}{27}\zeta_3
-\frac{2579}{27}\zeta_2
-7 
  \biggr)
  +\CS{C_A C_F^2 \nf}\biggl(  \frac{176}{9}\zeta_3
-\frac{1859}{27} 
  \biggr)
  +\CS{C_F^3 \nf}\biggl(  23 
  \biggr)
\nonumber \\ &
  +\CS{C_A^2 C_F \nf}\biggl(  \frac{280}{9}\zeta_6
+232\zeta_3^2
-80\zeta_5
-8\zeta_3 \zeta_2
+\frac{65}{3}\zeta_4
-\frac{2182}{9}\zeta_3
+\frac{34}{3}\zeta_2
-\frac{22627}{486} 
  \biggr)
\nonumber \\ &
  + \CS{\frac{d_A^{abcd}d_F^{abcd} \nf}{N_A}}\biggl(  \frac{296}{9}\zeta_6
+256\zeta_3^2
-\frac{4880}{3}\zeta_5
+608\zeta_3 \zeta_2
+\frac{664}{3}\zeta_4
+\frac{320}{3}\zeta_3
-160\zeta_2
+\frac{224}{9} 
  \biggr)
\nonumber \\ &
  +\CS{C_A^2 \nf^2}\biggl(  -\frac{8}{9}\zeta_5
-\frac{32}{9}\zeta_3 \zeta_2
+\frac{200}{27}\zeta_4
+\frac{289}{27}\zeta_3
+\frac{37}{27}\zeta_2
+\frac{1352}{81} 
  \biggr)
  +\CS{C_A C_F \nf^2}\biggl(  \frac{160}{9}\zeta_3
+\frac{3910}{243} 
  \biggr)
\nonumber \\ &
  +\CS{C_F^2 \nf^2}\biggl(  -\frac{176}{9}\zeta_3
+\frac{338}{27} 
  \biggr)
  +\CS{\frac{(d_F^{abcd})^2 \nf^2}{N_A}}\biggl(  \frac{512}{3}\zeta_3
-\frac{704}{9} 
  \biggr)
  +\CS{C_A \nf^3}\biggl(  \frac{5}{243} 
  \biggr)
  +\CS{C_F \nf^3}\biggl(  \frac{154}{243} 
  \biggr)
\end{align}
for the virtual anomalous dimension of the gluon.
$B_g^{(4)}$ was derived in this way also in \cite{Vogt:2026lltalk,Gardi:2026b4gemail,Moch:2026qsw} and we agree with their result.

\section{Conclusions}
We presented a complete calculation of the four-loop, non-singlet splitting functions in QCD.
In these proceedings, we discussed how this result can be used to derive the eikonal soft anomalous dimension and the virtual anomalous dimension for both quarks and gluons.
It is expected that the computational method employed here can also be used to calculate the singlet splitting function.
The latter, however, requires additional ingredients for the renormalization due to the impact of gauge-variant operators.

\acknowledgments

We are grateful to Johannes Bl\"umlein, Einan Gardi and Andreas Vogt for enlightening discussions and comparisons.
We thank Peter Marquard and Matthias Steinhauser for their exceptional efforts in hosting a successful, inspiring, and pleasant 2026 edition of the {\em Loops and Legs in Quantum Field Theory} conference series.
This work has been supported by the European Research Council (ERC) 
under the European Union's Horizon 2020 research and innovation programme grant agreement 101019620 (ERC Advanced Grant TOPUP).
\bibliographystyle{JHEP}
\bibliography{refs}

%

\end{document}